\documentclass{bmcart}
\usepackage{amsthm,amsmath}
\usepackage[utf8]{inputenc}
\usepackage{kotex}
\usepackage{graphicx}
\usepackage{hyperref}
\usepackage{CJKutf8}
\usepackage{graphicx}
\usepackage{amsmath, amssymb}
\usepackage{color, xcolor}
\usepackage{multirow}
\usepackage{array}
\usepackage{todonotes}
\usepackage{setspace}
\usepackage{pdflscape}
\usepackage{longtable}
\usepackage{caption}
\usepackage[title]{appendix}

\begin{document}

\begin{frontmatter}

\begin{fmbox}
\dochead{Research}

\title{Quantifying interdisciplinary synergy in higher STEM education}

\author[
   addressref={aff1},
   email={dotchgahyoun@gnu.ac.kr}
]{\inits{GG}\fnm{Gahyoun} \snm{Gim}}
\author[
   addressref={aff2}, 
   email={jinhyuk.yun@ssu.ac.kr}
]{\inits{JY}\fnm{Jinhyuk} \snm{Yun}}
\author[
   addressref={aff1,aff3,aff4}, 
   corref={aff1}, 
   email={lshlj82@gnu.ac.kr}
]{\inits{SHL}\fnm{Sang Hoon} \snm{Lee}}

\address[id=aff1]{%
  \orgname{Department of Physics, Gyeongsang National University}, 
  \street{Jinju}, 
  \postcode{52828}, 
  \cny{Korea}
}
\address[id=aff2]{%
  \orgname{School of AI convergence, Soongsil University}, 
  \street{Seoul}, 
  \postcode{06978}, 
  \cny{Korea}
}
\address[id=aff3]{%
  \orgname{Research Institute of Natural Science, Gyeongsang National University}, 
  \street{Jinju}, 
  \postcode{52828}, 
  \cny{Korea}
}
\address[id=aff4]{%
  \orgname{Future Convergence Technology Research Institute, Gyeongsang National University}, 
  \street{Jinju}, 
  \postcode{52849}, 
  \cny{Korea}
}

\end{fmbox}

\begin{abstractbox}
\begin{abstract}
We propose a framework to quantify and utilize interdisciplinarity in science
and engineering curricula at the university-level higher education.
We analyze interdisciplinary relations by standardizing large-scale official
educational data in Korea using a cutting-edge large language model and
constructing knowledge maps for disciplines of scientific education. We
design and evaluate single-field and integrated dual-field curricula by
adapting pedagogical theory and utilizing information theory-based metrics.
We develop standard curricula for individual disciplines and integrated
curricula combining two fields, with their interdisciplinarity quantified
by the curriculum synergy score. The results indicate higher interdisciplinarity
for combinations within or across closely related fields, especially in
engineering fields. Based on the analysis, engineering fields constitute
the core structure of our design for curriculum interdisciplinarity, while
basic natural science fields are located at peripheral stems to provide
fundamental concepts.
\end{abstract}
\end{abstractbox}
\end{frontmatter}

\section{Introduction}
\label{sec:intro}

The increasing complexity of challenges across the social, academic, and
industrial domains has focused on interdisciplinary approaches to solving
advanced problems~\cite{Ledford2015}. In response to this social demand,
higher education institutions, e.g., universities, are progressively introducing
and expanding their multidisciplinary programs~\cite{Tsao2024,Newell1998,Klein2006,Ashby2019}.
Consequently, researchers are actively working on the interdisciplinary
theory and its curriculum models nowadays~\cite{Tsao2024,Penprase2020}.
Unfortunately, most such research relies on qualitative assessments based on 
theoretical arguments or small-scale data, calling into question its generalizability.

The primary objective of this research is to establish a principled data-driven framework for designing and evaluating interdisciplinary-integrated curricula. To achieve this, we analyze large-scale real-world data extending beyond traditional curriculum design approaches. We develop a principled data processing pipeline and a quantitative framework to measure interdisciplinarity across large-scale academic curricula. Our methodology begins with standardizing extensive scientific and technological natural language data. We build a standardized dataset using cutting-edge natural language processing (NLP) techniques, such as large-scale language models (LLMs)~\cite{geminiteam2024gemini15unlockingmultimodal}, and higher-order methods~\cite{Boccaletti2023}. We link this dataset to departments across institutions through a bipartite network to construct knowledge maps. By projecting this network, we analyze standard subjects' relationships and interconnections among various disciplines.

In particular, we focus on the science, technology, engineering, and math (STEM) disciplines. The decision to concentrate on the STEM disciplines in our analysis is largely influenced by the dynamic evolution and heightened demand these disciplines are experiencing, driven by recent technological advancements. The STEM disciplines are at the forefront of innovation, playing a crucial role in addressing some of the world's most pressing challenges. As a result, there is an increasing demand for professionals skilled in these areas, necessitating a stronger emphasis on interdisciplinarity to equip future experts with the ability to integrate knowledge and skills from various domains~\cite{WashingtonPost}. Recent policy directions, including those led by the U.S. National Science Foundation, emphasize the necessity of interdisciplinary training in STEM to prepare the next generation of leaders for complex, real-world challenges~\cite{NSF2024InterdisciplinaryPolicy}. New policies are advocating for an interdisciplinary approach within STEM education, recognizing that the complexity of modern challenges requires solutions that are not confined to a single discipline~\cite{NRC}. By focusing on STEM, this research aims to contribute to the development of educational frameworks that not only respond to these policy shifts but also prepare students to thrive in a rapidly evolving technological landscape.

Leveraging information theory~\cite{Cover2006} to show the difference between the major distributions quantitatively, we propose a single-discipline standard curriculum model that reflects real-world constraints such as time, resources, and institutional policies. In addition, we hypothesize integration majors, which combine two distinct fields, and develop a double-field integrated curriculum for each possible combination. To assess the interdisciplinarity of these integration majors, we present a measurement framework 
taking both the curriculum similarity and the synergy into account. This metric allows us to determine whether a given double-field integrated curriculum achieves true interdisciplinarity or merely a multidisciplinary combination by systematically evaluating all possible combinations and comprehensively analyzing their degree of integration.

The paper is organized as follows. In Sect.~\ref{sec:data_processing}, we introduce our data and the process of course standardization with the large language model (LLM). Based on the processed data, we first analyze the structural properties of the department similarity in terms of curricula in Sect.~\ref{sec:DSN}. As the main topic of our work, we present the interdisciplinary curriculum model and the quantitative measure of synergy between disciplines in Sect.~\ref{sec:curriculum_design}. We summarize our work and discuss its implications in Sect.~\ref{sec:conclusion}.

\section{Data processing of course standardization}
\label{sec:data_processing}

For this research, we develop a data processing framework that allows us to assess the quality of intermediate steps and final results across sectors and scales. To present higher education boundary's standard single-field curricula and double-field integrated curricula and quantify their multidisciplinarity and interdisciplinarity, we employ datasets primarily sourced from the  ``Higher Education in Korea'' portal publicly operated by the Ministry of Education in Korea for the 2024 spring semester. 
The data sets are officially collected, processed, and validated by the Ministry of Education in Korea as a national census of higher education, so it is highly authoritative and trustworthy.
Our primary data set includes the  ``Curriculum per Educational Unit by School'' for the 2024 spring semester~\cite{Curriculum2024}. In addition, we utilize the  ``List of Undergraduate Departments and Majors by School'' for the same semester~\cite{Departments2024}, as metadata to identify departments and educational fields. These data sets were integrated to extract STEM departments' curricula from departments classified under the  ``Standard Classification Information (표준분류정보 in Korean)'' system~\cite{data_go_kr}.

As introduced in Sect.~\ref{sec:intro}, we focus on the departments running four-year undergraduate programs within STEM fields classified under Engineering and Natural Sciences in the Standard Classification (the highest-level classification). Specifically, within Natural Sciences, we extracted data from departments categorized as ``Basic Natural Sciences,'' which include ``Chemistry, Life Science, and Environment'' and ``Mathematics, Physics, Astronomy, and Earth Science'' in the Standard Classification (the intermediate-level classification). As a result of our preprocessing and comprehensive data integration, we constructed a standardized dataset comprising curricula from $2\,841$ STEM departments\footnote{The original data included $2\,851$ unique STEM undergraduate departments. However, due to military security restrictions and data structure errors, only $2\,841$ departments were ultimately analyzed.} across $161$ higher education institutions, including a total of $126\,437$ raw course entries. We use departments as the primary unit of analysis because our data, sourced from the Ministry of Education in Korea, is organized at this level, providing a consistent framework with around $80$ credits (for major courses) over four years per department. In Korea, departments serve as the \emph{de facto} units of disciplines due to the structured nature of higher education, aligning with current educational practices and government initiatives that encourage interdisciplinary integration. While finer distinctions at the program level might be desirable, current data limitations necessitate this approach. Future research could explore more granular analyses as detailed data becomes available, such as through ``microdegree'' programs aimed at fostering small units of interdisciplinary education. 

We also note that different departments may offer courses with varying credit allocations, but the overall credit requirement standardizes the instructional volume, facilitating consistent cross-departmental comparisons. All curricula are assessed under a rather uniform constraint in the total number of credits, ensuring that variations in course numbers do not bias our analysis. This approach is again consistent with Korea's centralized higher education policy, characterized by a national standard for four-year undergraduate curricula. Although such discrepancies might have greater implications in countries with more flexible curricula, their impact would be minimal in the Korean educational context. 

The courses are obviously building blocks of the coursework in each department. However, we cannot simply use the raw course data as it is, because there are multiple nomenclatural variants of essentially the same course, e.g., ``quantum mechanics'' and ``quantum physics'' in physics-related departments. Therefore, it is crucial to \textit{standardize} different variants of course names in the raw data to accurately identify course units that are as meaningfully independent as possible shared by different departments and institutions. Without the process, only the courses with exactly the same name would be identified as the same entity. Such a task in this scale would be next to impossible manually, but cutting-edge LLMs enable us to do this task systematically. In particular, we use Gemini-1.5 Pro~\cite{geminiteam2024gemini15unlockingmultimodal}, which is commonly referred to as one of the most accurate language models as of July 2024, when we performed this part of data processing.  
By employing higher-order data processing techniques combined with the LLM, which will be discussed in Sect.~\ref{sec:highorder_processing}, we systematically process these raw courses into $24\,476$ standardized courses. 
Our approach enables us to effectively process large-scale datasets while accurately capturing the semantic complexities inherent in natural language and disciplinary knowledge.

\begin{figure}
\includegraphics[width=0.95\textwidth]{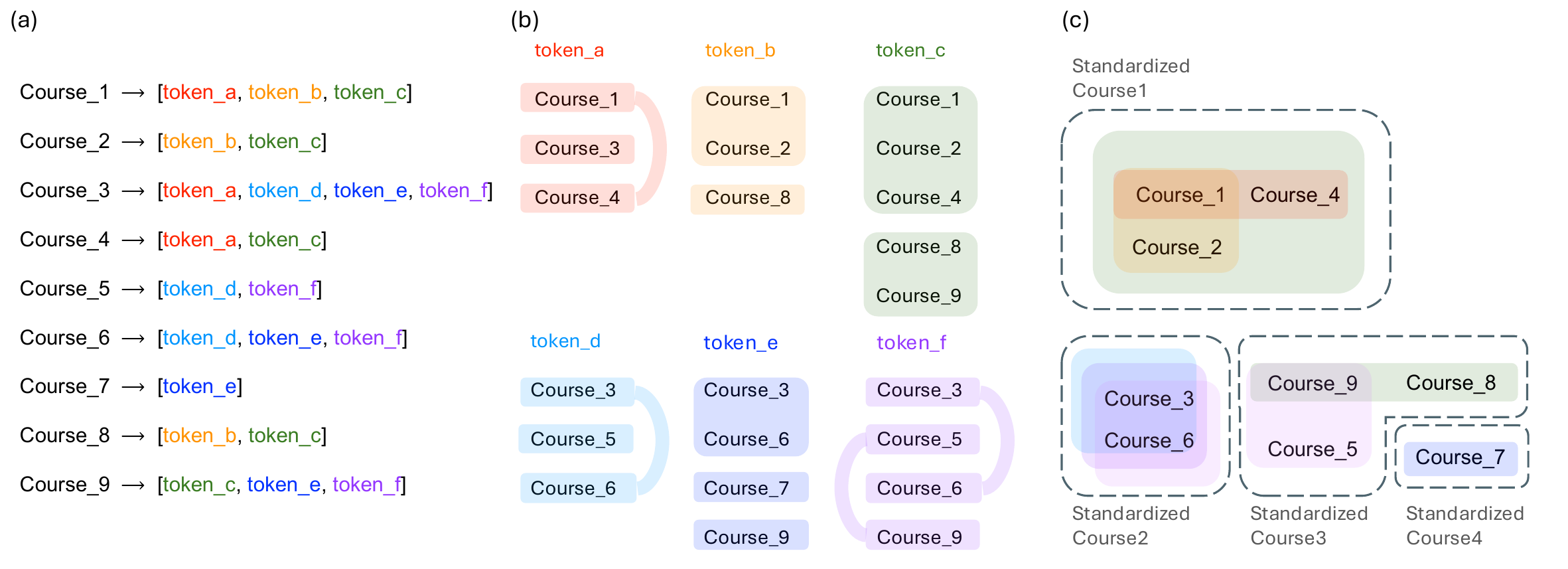} 
\caption{A schematic illustration of education data processing with LLM. (a) The tokenizing process of course data by LLM (Sect.~\ref{sec:tokenizing}). Each course is broken into distinct tokens, marked by different colors. (b) For each token, the courses including the token are listed below. Among them, the courses identified as the same by the LLM are connected in the form of hyperedges (enclosed by the same connected shaded area). 
(c) Each connected component constructed by collecting the hyperedges in the process described in the panel (b) is identified as each standardized course.}
\label{figure: llm and higher-order processing}
\end{figure}

\subsection{Tokenizing from the raw course data with LLM}
\label{sec:tokenizing}

Even with the LLM, extracting the standardized courses directly from the exhaustive list of $126\,437$ raw courses is not plausible due to the combinatorial explosion. Our strategy to circumvent this problem to proceed is the following. We first utilize tokenization to segment each of the raw course names into meaningful components that serve as the basic indicators for further processing. Each course \( C_i \) is decomposed into a set of tokens $\{t_1, t_2, \dots, t_m\}$ as illustrated in Fig.~\ref{figure: llm and higher-order processing}(a). 
To ensure accurate processing of the natural language in the raw course names (Korean), we use prompts tailored to its linguistic properties with specific examples (prompt~\ref{prompt1} in Appendix~\ref{sec:prompts}).
Through this tokenization process, we obtain $8\,841$ unique tokens from the raw course data. 
By focusing on the shared tokens among different courses, this preprocesses the data to highlight potential semantic similarities for further analysis. To process raw courses effectively, for each token, we collect the list of courses including it as illustrated in Fig.~\ref{figure: llm and higher-order processing}(b). 

\subsection{Constructing hyperedges among raw courses with LLM}
\label{sec:highorder_processing}

Now we have the set of raw courses for each token $t$ as in Fig.~\ref{figure: llm and higher-order processing}(b), and our next step is to collect the effectively same raw courses as intermediate-level units. Mathematically, this step involves identifying higher-order relationships (i.e., hyperedges)~\cite{Boccaletti2023} among raw courses---beyond simple pairwise associations---so as to capture more nuanced semantic groupings that reflect disciplinary structures~\cite{Aksoy2020, Kumar2020, Citraro2023}.
Similar to the tokenization process described in Sect.~\ref{sec:tokenizing}, we again use a prompt that effectively clusters identical courses based on shared topics or subfields (prompt~\ref{prompt2} in Appendix~\ref{sec:prompts}), depicted in Fig.~\ref{figure: llm and higher-order processing}(b). 

While conventional NLP techniques have long been applied to large-scale textual data analysis, they often rely on rigid pipelines tailored to narrow tasks. In contrast, we adopt large language models (LLMs) due to their superior contextual understanding and adaptability. Recent studies have demonstrated that LLMs consistently outperform traditional NLP tools in extracting nuanced semantic relationships from domain-specific scientific corpora~\cite{LLMOutperformsNLP2025}. Their ability to integrate context, handle ambiguous terminology, and operate with minimal labeled data makes them especially effective for tasks involving the clustering of STEM course descriptions, which are rich in discipline-specific language and complex structural patterns.

Through this LLM-based higher-order clustering process, we generate $91\,964$ hyperedges from all tokens, which encapsulate the intermediate-level interrelationships present in the raw course data.
\subsection{Standardizing courses}
\label{sec:Standardizing_Courses}

Our final step to standardize courses is based on the list of hyperedges in Sect.~\ref{sec:highorder_processing}. As illustrated in Fig.~\ref{figure: llm and higher-order processing}(c), we merge all of the courses inside each of the connected components.
Intuitively, this definition captures not only directly connected courses within a hyperedge but also courses that are indirectly linked through the chain of associations. For example, if course \( C_1 \) is associated with hyperedge \( H_k \) and course \( C_2 \) is linked to both \( H_k \) and \( H_l \), then \( C_1 \) and \( C_2 \) belong to the same connected component.
The set of connected components constitutes our final set of standardized courses \( \mathcal{S} \).
In this way, each connected component groups the raw courses that are semantically or structurally similar based on shared tokens or associations identified via the hyperedge construction process.

Through this standardization process, we reduce $126\,347$ raw courses to $24\,476$ standardized courses from STEM departments of Korean universities. This reduction merges redundant or overlapping courses into unified representations while preserving their semantic integrity and higher-order relationships, for more interpretable and robust analysis.
By our processing, the standardized courses provide a deeper understanding of the underlying structures and interconnections within curricula~\cite{Lemaire2006}.  
For a concrete example, we provide the step-by-step process in the case of statistical physics in 
Appendix~\ref{sec:example}. 

\section{Department similarity based on standardized courses}
\label{sec:DSN}

\begin{figure}
\centering 
\includegraphics[width=0.95\textwidth]{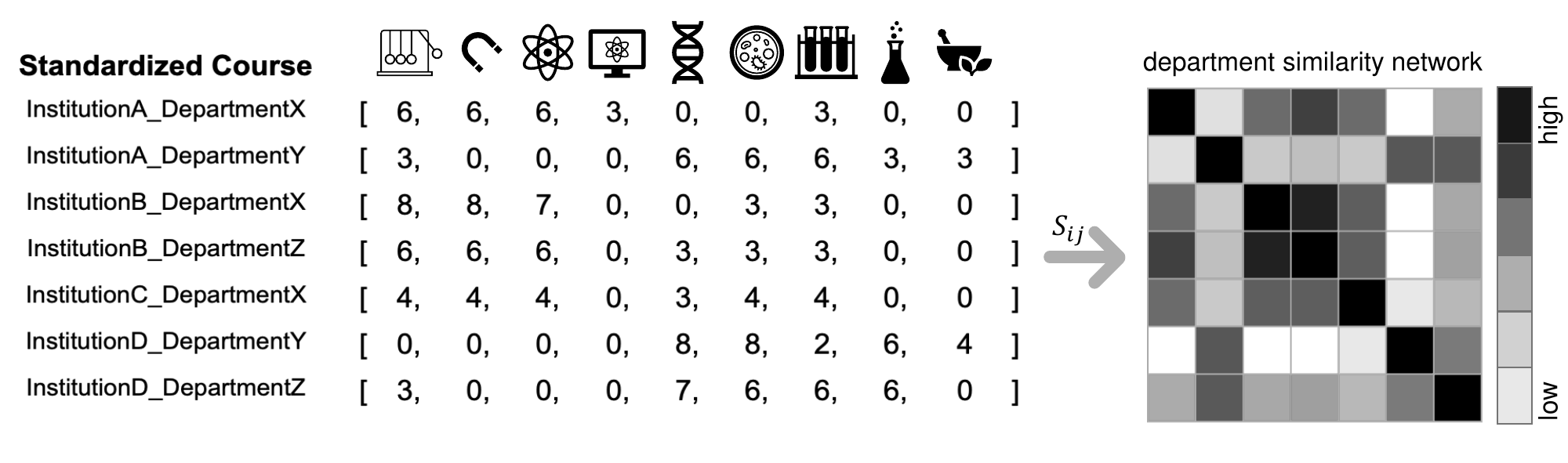}
\caption{A schematic illustration of the DSN constructing process based on standardized course data.  
The vectors on the left represent the credits for each standardized course for multiple departments across institutions, where each element corresponds to the total weight (the credit sum) assigned to a specific standardized course. Each row represents a department, and each column corresponds to a standardized subject.  
The right matrix represents the pairwise similarity \( S_{ij} \) according to Eq.~\eqref{eq:S}.  
The resulting similarity matrix is visualized as a heatmap, where darker shades indicate higher similarity between departments.}
\label{figure: Schematic viz of processing-1}
\end{figure}

\begin{figure}
\centering
\includegraphics[width=0.95\textwidth]{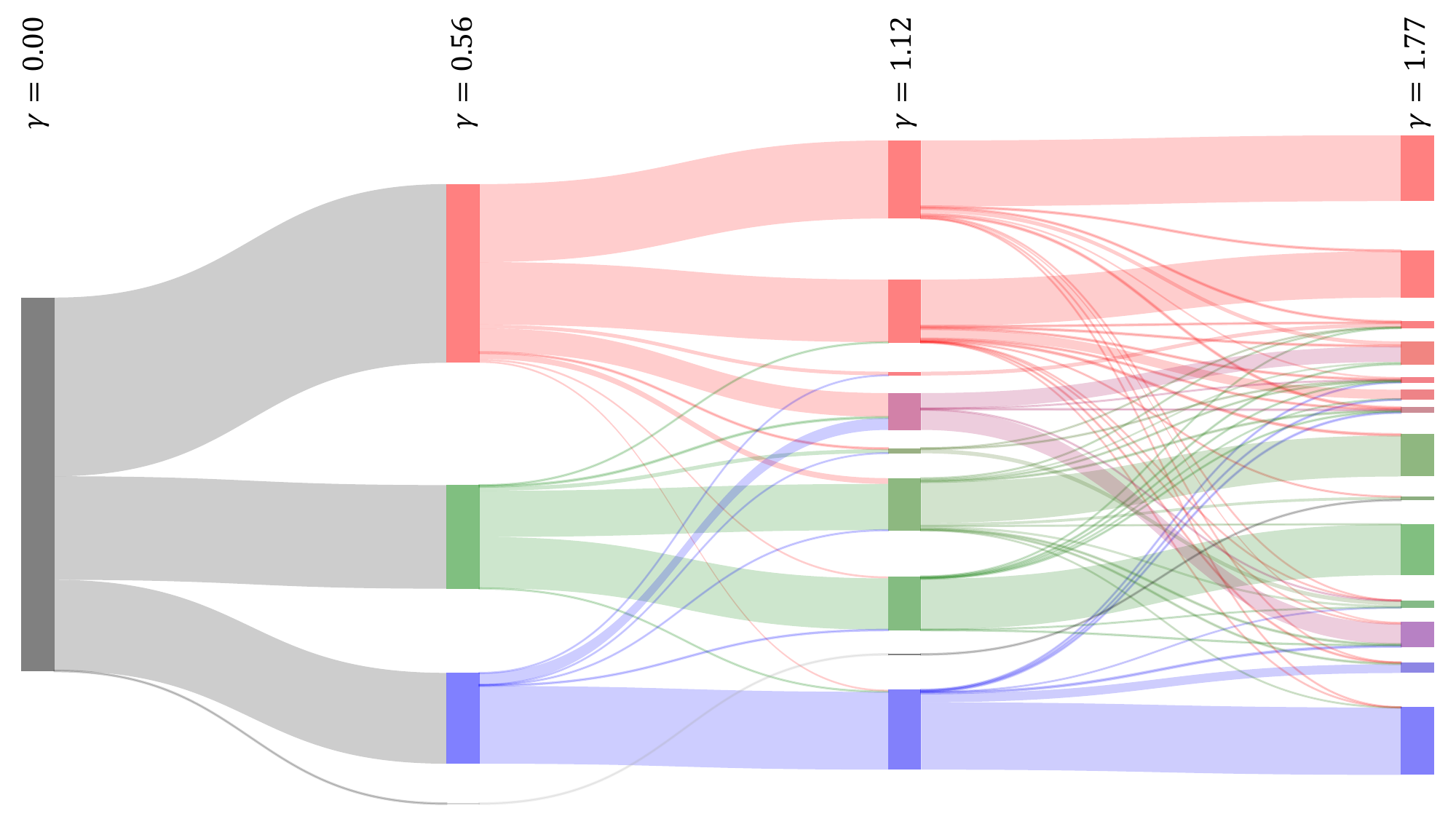} 
\caption{Alluvial diagram showing the membership reorganization of community structures at four representative resolution values: $\gamma = 0$, $0.56$, $1.12$, and $1.77$, where the community structures show global consistency (see Appendix~\ref{sec:PaI} for details), 
where the vertical positions of the nodes are determined to minimize crossing.
At $\gamma = 0.56$, each node and link represents the community's aggregated departmental leanings (PSE: red, TSE: green, and MLSE: blue), and at finer resolution values, the weighted sum of colors represents the reorganization of departmental classifications.}
\label{figure: department similarity network}
\end{figure}

To verify the validity and utility of the standardized courses extracted in Sect.~\ref{sec:data_processing}, we constructed a department similarity network (DSN), where nodes represent departments, and the weighted edges represent their department similarity in terms of their curricula (Fig.~\ref{figure: Schematic viz of processing-1})~\cite{NewmanBook,WeightedNetwork}. The department similarity is quantified using the weighted Jaccard similarity coefficient, defined as:
\begin{equation}
S_{ij} = \frac{\displaystyle \sum_{k=1}^{n} \min(c_{ik}, c_{jk})}{\displaystyle \sum_{k=1}^{n} \max(c_{ik}, c_{jk})},
\label{eq:S}
\end{equation}
where \( c_{ik} \) represents the credit-weighted value of the standardized course \( k \) (the credit of a standardized course is defined as the sum of the credits assigned to the raw courses corresponding to the standardized course) in department \( i \), and \( S_{ij} \) measures the proportion of standardized coursework shared between departments \( i \) and \( j \). 

We explore the structural properties of DSN from the perspective of communities~\cite{Porter2009,Fortunato2010} of various scales. We take the popular modularity-maximization algorithm Louvain~\cite{Blondel2008} with the tunable resolution parameter $\gamma$. We choose four representative values of $\gamma = 0$, $0.56$, $1.12$, and $1.77$, where the overall community structures show global consistency among the results (Appendix~\ref{sec:PaI})~\cite{PhysRevE.103.052306}. We illustrate the scale-dependent reorganization of communities across the four resolution values as an alluvial diagram in Fig.~\ref{figure: department similarity network}.
Each block in the diagram corresponds to a community within DSN at a particular $\gamma$ value, with the inter-connection between the blocks in the adjacent $\gamma$ values indicating the flow of department membership. 

At \( \gamma = 0 \), the DSN consists of a single community by the mathematical property of the modularity function~\cite{Porter2009,Fortunato2010}, marked by gray. As the resolution increases to \( \gamma = 0.56 \), the network is divided into three main distinct communities: (1) pattern science and related engineering (PSE) such as physics, mathematics, and software engineering (the top red block), (2) technical systems engineering (TSE) such as architectural engineering (the middle green block), and (3) molecular, life science, and related engineering (MLSE) such as chemistry, biology, and chemical engineering (the bottom blue block). Departments that do not belong to these main clusters correspond to the thin gray block. It is interesting to note that the first meaningful separation of communities is not the division between natural science and engineering that one might imagine, but that between PSE and MLSE, where different types of sciences mixed with various engineering departments for both, in addition to TSE. 

As we increase the resolution parameter further, at an intermediate resolution value \( \gamma = 1.12 \) and the finest resolution \( \gamma = 1.77 \), the aforementioned three communities are refined further. The colors of the communities at these resolutions are adjusted to reflect their original affiliations at \( \gamma = 0.56 \). When departments from different communities are remerged into a single community at a finer resolution value, their colors are blended to highlight the interdisciplinary composition. Note that as the resolution increases, the red (PSE) and green (TSE) communities are fragmented into smaller communities. In contrast, the blue (MLSE) communities retain a significant portion of their original nodes, reflected by similar block sizes even at finer resolution values.

One can track the membership reorganization of specific departments across different scales, and it would be interesting to see the flow of less traditional and interdisciplinary departments. 
For example, in the case of the \textit{Department of Scientific Computing} at Pukyong National University, whose curriculum is highly interdisciplinarity with the studies on scientific computing from various disciplines, belongs to a computer-science-related community at intermediate resolution values. However, as the resolution gets finer, this department transitions into the physics-related community presumably because of its interdisciplinary curriculum.

\section{Interdisciplinary curriculum design and quantifying its synergy}
\label{sec:curriculum_design}

\begin{figure}
\centering 
\includegraphics[width=0.95\textwidth]{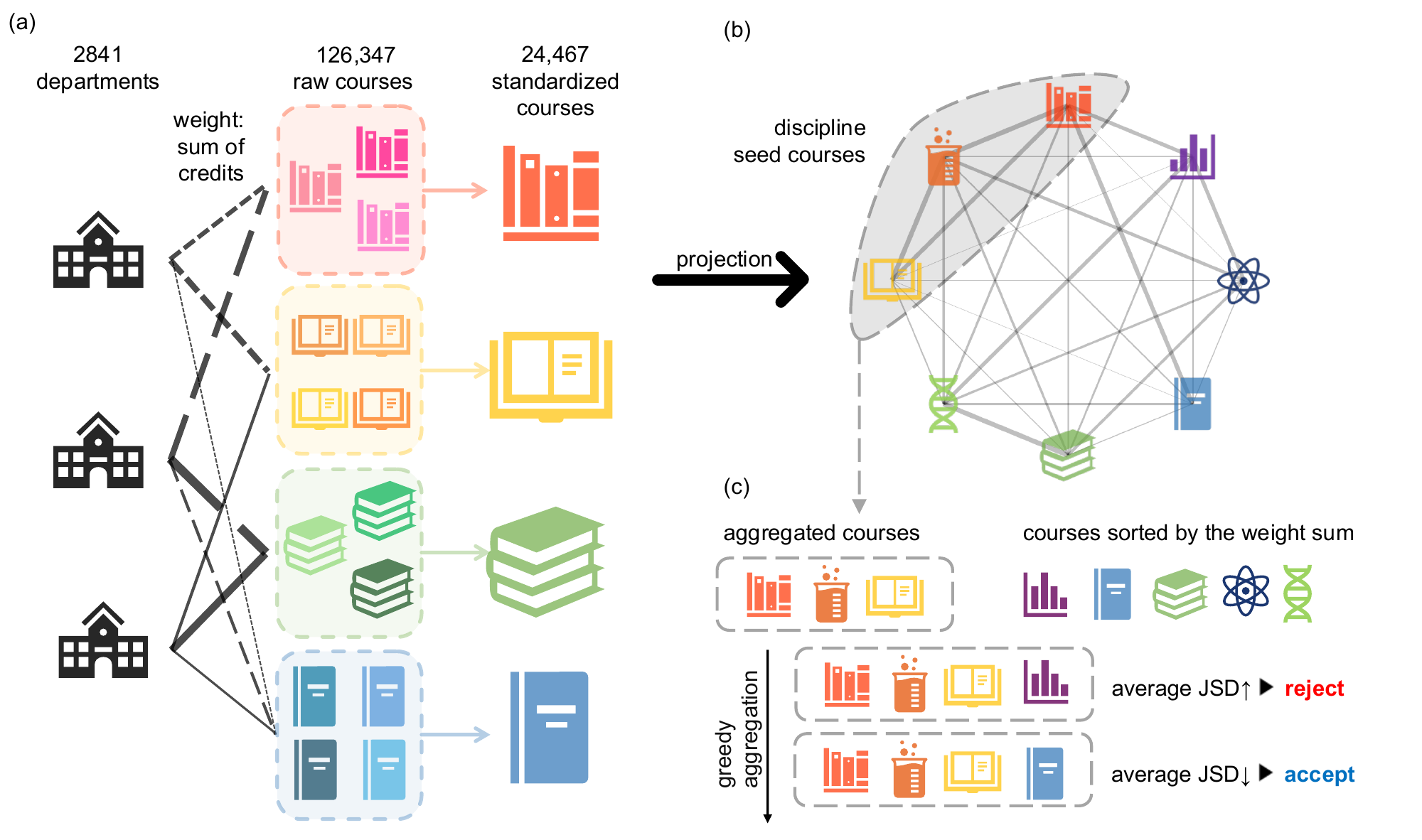}
\caption{A schematic diagram of our curriculum design process. 
(a) A recap of the weighted bipartite network between departments and standardized courses, described as a weighted matrix in Fig.~\ref{figure: Schematic viz of processing-1}. The detailed course standardization process is stated in Sect.~\ref{sec:data_processing}. 
(b) The standardized-course-mode projection from the bipartite network, from which the curriculum is aggregated by sequential addition of candidate courses based on the information-theoretical criterion described in panel (c). 
}
\label{figure: Schematic viz of processing}
\end{figure}

In this section, based on the standardized courses, we suggest a framework for designing curricula for potential interdisciplinary fields, followed by evaluating the synergy from such field combinations. First, we need to set up a list of standard academic disciplines.
As a reminder, we have constructed a comprehensive list of STEM curricula in terms of standardized courses for each department in Sect.~\ref{sec:data_processing} and Sect.~\ref{sec:DSN}. This process involves constructing a bipartite network with nodes representing $24\,467$ standardized courses and $2\,841$ STEM departments. The edges in the bipartite network are weighted by the total credit hours associated with each course [Figs.~\ref{figure: Schematic viz of processing-1} and \ref{figure: Schematic viz of processing}(a)]. To quantify the relationship between the standardized courses, we project the bipartite network into the unipartite network composed of standardized courses, whose edges are weighted by the sum of credits in the departments sharing the course pairs [Fig.~\ref{figure: Schematic viz of processing}(b)].

\subsection{Single-discipline curriculum model}
\label{sec:single_discipline_curriculum}

From the original source data provided by Ministry of Education in Korea~\cite{data_go_kr}, the classification of ``Standard Classification information (the lowest-level classification)'' is designated as the list of $51$ independent disciplines\footnote{Among the $52$ disciplines in the data set~\cite{data_go_kr}, we exclude the ``not classified elsewhere (N.C.E.)'' entry for proper analysis. 
}.
For each discipline, we define the standardized credits of the standardized courses in the discipline as the most frequent nonzero credits in the departments belonging to the discipline. For instance, if the credits of a standardized course $\mathcal{C}$ appear as $\{ 6, 6, 3, 6, 4, 0, 0, 0, 0, 0 \}$ in $10$ departments corresponding to a discipline $\mathcal{D}$, the standardized credit of $\mathcal{C}$ in discipline $\mathcal{D}$ is $6$. From now on, ``courses'' and ``credits'' always refer to standardized ones, unless stated otherwise.

First, we need to design a curriculum model for a single discipline before moving on to the interdisciplinary curriculum. 
To design a standard single-field major curriculum for each discipline, we begin with five ``seed'' courses as the foundation, by choosing the five courses with the maximum credit assigned to the discipline\footnote{In the case of ties, we choose a desired number of courses uniformly at random---the missing courses by chance are likely to be included at the initial stage of model, so it does not make meaningful difference.}. The ``aggregation'' process of course attachment is composed of the two parts: sorting the candidate courses in the descending order of the weight sum between each course with the currently existing courses in the course-projected network [Fig.~\ref{figure: Schematic viz of processing}(b)], deciding whether we accept the candidate course (from the interactively updated queue of the sorted list) in the curriculum or not, and repeating this trial until we reach a certain threshold. 
A state of a curriculum during the aggregation process is described by the $n$($=|\mathcal{S}|$)-dimensional vector
\( \mathbf{w}_\mathrm{model} = \left( w_1, w_2, \cdots, w_n \right) \), where each index corresponds to a course and its corresponding element \( w_i \) represents the course \( i \)'s credit in the discipline.

To evaluate the degree of new information introduced by the interdisciplinary curriculum, we compute the Jensen-Shannon divergence (JSD) between the course-weighted vectors of the simple union and the integrated model. We denote this divergence as \( \mathcal{J} \), which provides a symmetric and bounded measure of dissimilarity between the two distributions:
\begin{equation}
\mathcal{J}\left(\mathbf{w}_\mathrm{model},\mathbf{w}_d \right) = \frac{1}{2} \left[ D_\text{KL}\left(\mathbf{w}_\mathrm{model} \middle\| \mathbf{m}_d \right) 
    + D_\text{KL}\left(\mathbf{w}_d \middle\| \mathbf{m}_d \right) \right] \,,
\end{equation}
where \( \mathbf{m}_d \) is the element-wise average vector of the two vectors:
\begin{equation}
    \mathbf{m}_d = \frac{1}{2} \left( \mathbf{w}_\mathrm{model} + \mathbf{w}_d \right) \,,
\end{equation}
and the Kullback-Leibler (KL) divergence \( D_\text{KL} \) measures the directed discrepancy between a given weight vector \( \mathbf{w} \) and a reference vector \( \mathbf{m} \),
\begin{equation}
D_\text{KL}\left(\mathbf{w} \middle\| \mathbf{m}\right) = \sum_{i=1}^{n} w_i \log \left( \frac{w_i}{m_i} \right) \,.
\label{eq:KL}
\end{equation} 
The process sequentially adds candidate courses similar to existing department curricula in the discipline, on average. At each step, the single-field average JSD is calculated as:
\begin{equation}
\langle \mathcal{J} \rangle_\mathcal{D} = \frac{1}{|\mathcal{D}|} \sum_{d \in \mathcal{D}} \mathcal{J}\left(\mathbf{w}_\mathrm{model},\mathbf{w}_d \right) \,.
\label{eq:JSD_D}
\end{equation} 
The addition of the candidate course in the model is accepted if $\langle \mathcal{J} \rangle_\mathcal{D} \le 0$, and rejected if $\langle \mathcal{J} \rangle_\mathcal{D} > 0$. Finally, we terminate the model aggregation when the sum of the credits reaches at least $80$, which is a typical number of credits for major courses in four-year undergraduate degrees in Korea. The optimized curriculum of discipline $\mathcal{D}$, denoted by $\Omega \left( \mathcal{D} \right)$, can be expressed as the credit-weighted vector for each course
\begin{equation}
\tilde{\mathbf{w}}_\mathcal{D} = \left( \tilde{w}_{\mathcal{D};1}, \tilde{w}_{\mathcal{D};2}, \cdots,  \tilde{w}_{\mathcal{D};n} \right) \,,
\label{eq:single_opt_weight_vector}
\end{equation}
where $\tilde{w}_{\mathcal{D};i}$ is the credit of course $i$ in the optimized curriculum of discipline $\mathcal{D}$. The vector for each discipline is standardized by the optimization process described in Sect.~\ref{sec:data_processing}, resulting in the uniform credit sum $\approx 80$, to ensure comparability and consistency in the curriculum design process.

\subsection{Interdisciplinary curriculum model}
\label{sec:interdisciplinary_curriculum}

To construct a double-field interdisciplinary curriculum model, we start with a seed of five courses in each field and their assigned credits for the $10$ courses in total. 
Similar to the single-discipline model in Sect.~\ref{sec:single_discipline_curriculum}, the candidate courses for addition are interactively sorted by the sum of weights [in the projected network in Fig.~\ref{figure: Schematic viz of processing}(b)] attached to the current courses in the model. 

In this case, the JSD for the acceptance criterion becomes
\begin{align}
\langle \mathcal{J} \rangle_{\mathcal{D}_A \otimes \mathcal{D}_B}  = 
\frac{1}{|\mathcal{D}_A|} \sum_{d_A \in \mathcal{D}_A} \mathcal{J}(\mathbf{w}_\mathrm{model}, \mathbf{w}_{d_A}) + 
\frac{1}{|\mathcal{D}_B|} \sum_{d_B \in \mathcal{D}_B} \mathcal{J}(\mathbf{w}_\mathrm{model}, \mathbf{w}_{d_B}) \,,
\end{align}
where \( \mathcal{D}_A \) and \( \mathcal{D}_B \) are  
the two respective disciplines.

This iterative process retains only the courses that reduce the average JSD across all of the departments belonging to both disciplines, and the same threshold of $80$ credits applies for the model termination. Similar to the single-discipline case, the optimized interdisciplinary curriculum of disciplines $\mathcal{D}_A$ and $\mathcal{D}_B$, denoted by $\Omega \left( \mathcal{D}_A \otimes \mathcal{D}_B \right)$, can be expressed as the credit-weighted vector for each course
\begin{equation}
\tilde{\mathbf{w}}_{\mathcal{D}_A \otimes \mathcal{D}_B } = \left( \tilde{w}_{\mathcal{D}_A \otimes \mathcal{D}_B;1}, \tilde{w}_{\mathcal{D}_A \otimes \mathcal{D}_B;2}, \cdots,  \tilde{w}_{\mathcal{D}_A \otimes \mathcal{D}_B;n} \right) \,,
\label{eq:interdisciplinary_opt_weight_vector}
\end{equation}
where $\tilde{w}_{\mathcal{D}_A \otimes \mathcal{D}_B;i}$ is the credit of course $i$ in the optimized interdisciplinary curriculum of disciplines $\mathcal{D}_A$ and $\mathcal{D}_B$.

\subsection{Quantifying interdisciplinarity with the curriculum synergy score}

In Sect.~\ref{sec:interdisciplinary_curriculum}, we have proposed an information-theoretical model to evaluate the degree of integration achieved by interdisciplinary curricula. The central aim is to design a curriculum that reflects both semantic cohesion across departments and pedagogical completeness within a combined field of study. To formalize this, we define an optimized curriculum set \( \Omega(\mathcal{D}) \) for any given set of departments \( \mathcal{D} \). The selection process is based on a semantic filtering criterion: each course must contribute to reducing the average divergence in Eq.~\eqref{eq:JSD_D} across departmental curricula. The credit constraint mentioned in Sect.~\ref{sec:single_discipline_curriculum} ensures that the resulting curriculum maintains the completeness required for a full major according to national education standards. An analogous formulation is applied to interdisciplinary combinations \( \mathcal{D}_A \otimes \mathcal{D}_B \).
Based on this formal construction, we introduce the curriculum synergy score (CSS), which evaluates whether the algorithmically constructed interdisciplinary curriculum \( \Omega \left( \mathcal{D}_A \otimes \mathcal{D}_B \right) \) achieves a deeper level of integration than a simple combination of the two disciplines, denoted by \( \Omega \left( \mathcal{D}_A \cup \mathcal{D}_B \right) \).

Although the concept of synergy is nontrivial, our operationalization focuses on two key aspects: (1) the interdisciplinary curriculum should maintain a sufficient degree of alignment with its constituent disciplines; and (2) it should generate new semantic content that is not trivially inherited from the union of its parts. This second aspect is grounded in the notion of ``surprisal" from information theory~\cite{Cover2006}, reflecting the emergence of integrative knowledge structures not found in either discipline alone. We assume the following course-weighted vector for the double-major,
\begin{equation}
\tilde{\mathbf{w}}_{\mathcal{D}_A \cup \mathcal{D}_B} = \left( \max\left\{ \tilde{w}_{\mathcal{D}_A;1}, \tilde{w}_{\mathcal{D}_B;1} \right\}, \cdots , \max\left\{ \tilde{w}_{\mathcal{D}_A;n}, \tilde{w}_{\mathcal{D}_B;n} \right\} \right) \,,
\label{eq:double_major_weighted_vector}
\end{equation}
which is simply the element-wise weighted union of the optimized curricula in Eq.~\eqref{eq:single_opt_weight_vector} for the two disciplines $\mathcal{D}_A$ and $\mathcal{D}_B$. 

Let us explore the mathematical interpretation of integrated disciplines through the lens of pedagogical theory. In contrast to multidisciplinarity, where disciplines coexist independently without significant integration, interdisciplinarity focuses on the meaningful synthesis of knowledge, tools, and perspectives from multiple fields~\cite{Wagner2011, Jacob2015}. Achieving interdisciplinary synergy requires a substantial overlap between disciplines, as opposed to merely combining them in a multidisciplinary manner. Our JSD-based CSS is specifically designed to capture this distinction by quantifying how effectively an integrated curriculum encompasses both shared similarities and the emergence of new information, as compared to a basic multidisciplinary curriculum. This approach helps distinguish truly integrative designs from mere aggregations by evaluating whether the synthesized curriculum provides added pedagogical value rather than duplicated content. For students, a high CSS implies a more coherent learning experience, minimizing redundancy and promoting cross-disciplinary connections that are more than the sum of their parts. To quantify the synergy reflecting those aspects, we employ an information-theoretic method based on prior research in educational fields~\cite{Klein2006, Ashby2019, Didham_Fujii_Torkar_2024, Choi2006, Jacobs2013, Stember1991}. For the new information emerging from the interdisciplinary curriculum, we again apply the JSD for the course-weighted vectors to quantify the informational distance between \( \Omega \left( \mathcal{D}_A \cup \mathcal{D}_B \right) \) and \( \Omega \left( \mathcal{D}_A \otimes \mathcal{D}_B \right) \), given by 
\begin{equation}
\mathcal{J}\left(\tilde{\mathbf{w}}_{\mathcal{D}_A \cup \mathcal{D}_B},\tilde{\mathbf{w}}_{\mathcal{D}_A \otimes \mathcal{D}_B} \right) = \frac{1}{2} \left[ D_\text{KL}\left(\tilde{\mathbf{w}}_{\mathcal{D}_A \cup \mathcal{D}_B} \middle\| \mathbf{m}_{\cup\otimes} \right) 
    + D_\text{KL}\left(\tilde{\mathbf{w}}_{\mathcal{D}_A \otimes \mathcal{D}_B} \middle\| \mathbf{m}_{\cup\otimes} \right) \right] \,,
\label{eq:finalJSD}
\end{equation}
where \( \mathbf{m}_{\cup\otimes} \) is the element-wise average vector of the two vectors:
\begin{equation}
    \mathbf{m}_{\cup\otimes} = \frac{1}{2} \left( \tilde{\mathbf{w}}_{\mathcal{D}_A \cup \mathcal{D}_B} + \tilde{\mathbf{w}}_{\mathcal{D}_A \otimes \mathcal{D}_B} \right) \,,
\end{equation}
and the KL divergence \( D_\text{KL} \) between two vectors is defined previously in Eq.~\eqref{eq:KL}. Note that in contrast to the JSD for designing the curriculum, the JSD here measures the amount of new information between \( \Omega \left( \mathcal{D}_A \cup \mathcal{D}_B \right) \) and \( \Omega \left( \mathcal{D}_A \otimes \mathcal{D}_B \right) \), so larger values indicate the possibility of unexpected synergy as a result of interdisciplinary curricula. Of course, as we previously mentioned, one cannot simply use the JSD alone because the ``unexpected'' information can be caused by synergetic or completely unrelated combinations. Therefore, we have to set an extra factor to filter those unrelated combinations. 

Our strategy for filtering is to use a similarity measure between the disciplines of interest. 
We use the following simple measure that quantifies the degree of shared courses between the two disciplines, measuring the extent of overlap in their optimized weighted course vectors. It is computed as
\begin{equation}
\sigma_{\mathcal{D}_A,\mathcal{D}_B} = \frac{\displaystyle \sum_{i=1}^n \min \left\{ \tilde{w}_{\mathcal{D}_A;i}, \tilde{w}_{\mathcal{D}_B;i} \right\}}{\displaystyle \sum_{i=1}^n \max \left\{ \tilde{w}_{\mathcal{D}_A;i}, \tilde{w}_{\mathcal{D}_B;i} \right\}} \,.
\label{eq:similarity}
\end{equation}
Finally, we define the CSS between disciplines $\mathcal{D}_A$ and $\mathcal{D}_B$ as the product of the two factors in Eq.~\eqref{eq:finalJSD} and Eq.~\eqref{eq:similarity},
\begin{equation}
\Theta \left( \mathcal{D}_A,\mathcal{D}_B \right) = \mathcal{J}\left(\tilde{\mathbf{w}}_{\mathcal{D}_A \cup \mathcal{D}_B},\tilde{\mathbf{w}}_{\mathcal{D}_A \otimes \mathcal{D}_B} \right) \sigma_{\mathcal{D}_A,\mathcal{D}_B} \,,
\label{eq:CSS}
\end{equation}
which takes both the new information from the interdisciplinary curriculum design and the discipline similarity into account. For instance, if two disciplines are too similar, $\sigma_{\mathcal{D}_A,\mathcal{D}_B} \approx 1$ but  $\mathcal{J}\left(\tilde{\mathbf{w}}_{\mathcal{D}_A \cup \mathcal{D}_B},\tilde{\mathbf{w}}_{\mathcal{D}_A \otimes \mathcal{D}_B} \right) \approx 0$ and thus $\Theta \left( \mathcal{D}_A,\mathcal{D}_B \right) \approx 0$. Another extreme case is the two totally unrelated disciplines, which may induce a large value of $\mathcal{J}\left(\tilde{\mathbf{w}}_{\mathcal{D}_A \cup \mathcal{D}_B},\tilde{\mathbf{w}}_{\mathcal{D}_A \otimes \mathcal{D}_B} \right)$ but $\sigma_{\mathcal{D}_A,\mathcal{D}_B} \approx 0$ and thus $\Theta \left( \mathcal{D}_A,\mathcal{D}_B \right) \approx 0$ again. An ideal interdisciplinary combination for synergy corresponds to a subtle balance between emergent information and decent similarity. 

\begin{figure}
\centering
\includegraphics[width=0.95\textwidth]{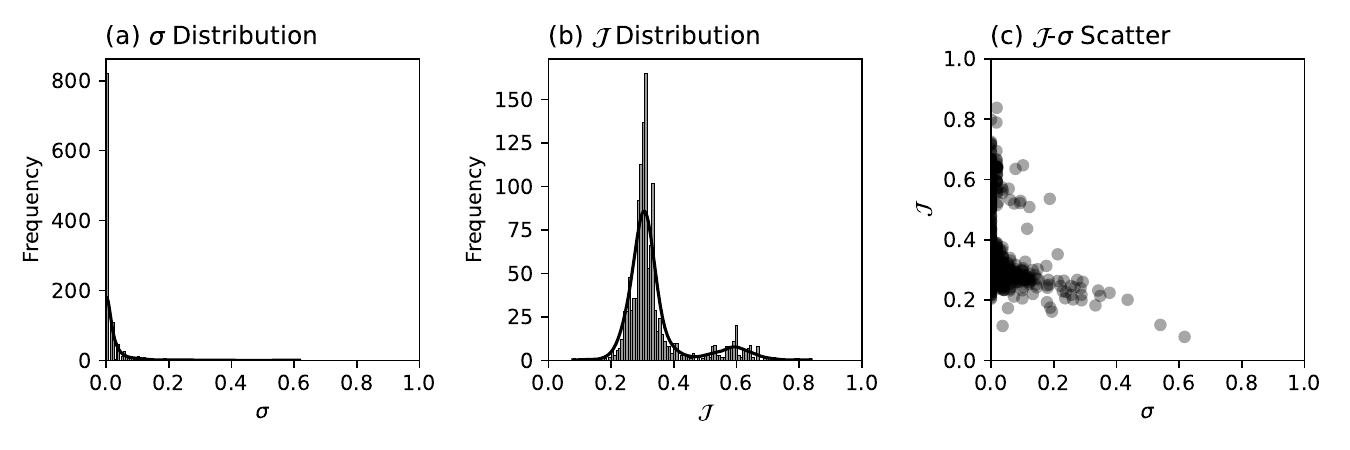} 
\caption{(a) The histogram of similarity $\sigma_{\mathcal{D}_A,\mathcal{D}_B}$ between discipline pairs. 
(b) The histogram of information distance $\mathcal{J}\left(\tilde{\mathbf{w}}_{\mathcal{D}_A \cup \mathcal{D}_B},\tilde{\mathbf{w}}_{\mathcal{D}_A \otimes \mathcal{D}_B} \right)$ between the double-major and interdisciplinary curriculum models for discipline pairs. 
(c) The scatter plot between them. 
}
\label{figure: histogram and scatter}
\end{figure}

The CSS, therefore, quantifies the synergy between disciplines that cannot solely be captured by simple similarity measures, which highlights the added value of proper interdisciplinary integration. 
We observe the distribution of each factor separately, and one can check that the vast majority of $\sigma_{\mathcal{D}_A,\mathcal{D}_B}$ values are close to $0$ [Fig.~\ref{figure: histogram and scatter}(a)], which is understandable by the fact that most discipline pairs would be quite different to each other. 
On the other hand, the distribution of $\mathcal{J}\left(\tilde{\mathbf{w}}_{\mathcal{D}_A \cup \mathcal{D}_B},\tilde{\mathbf{w}}_{\mathcal{D}_A \otimes \mathcal{D}_B} \right)$ shows the most prominent peak around $0.3$ presumably corresponding to the overall scale difference between $\tilde{\mathbf{w}}_{\mathcal{D}_A \cup \mathcal{D}_B}$ and $\tilde{\mathbf{w}}_{\mathcal{D}_A \otimes \mathcal{D}_B}$, and the second tallest peak around $0.6$ [Fig.~\ref{figure: histogram and scatter}(b)]. There is no notable correlation between $\sigma_{\mathcal{D}_A,\mathcal{D}_B}$ and $\mathcal{J}\left(\tilde{\mathbf{w}}_{\mathcal{D}_A \cup \mathcal{D}_B},\tilde{\mathbf{w}}_{\mathcal{D}_A \otimes \mathcal{D}_B} \right)$ [Fig.~\ref{figure: histogram and scatter}(c)], which also justifies the formalism of $\Theta \left( \mathcal{D}_A,\mathcal{D}_B \right)$ as the product of them. In Appendix~\ref{sec:loglog_sigma}, we provide the plots in the double-logarithmic scale in Fig.~\ref{fig:loglog_sigma} to help the reader better understand the range and tail behavior of the similarity values.

For notable examples, we provide three examples in supplementary tables: two examples of beneficial synergy (`physics' + `semiconductor engineering' in Table~\ref{table:physics_semiconductor_combination} and `chemistry' + `high polymer engineering' in Table~\ref{table:chemistry_high_polymer_combination}) and an example of no synergy (`physics' + `chemistry' in Table~\ref{table:physics_chemistry}).  Notably, science disciplines and their related engineering disciplines form better combinations than two science disciplines, and this tendency will be highlighted again in the analysis presented in the next subsection. Other examples include combinations `biotechnology + energy engineering,' which are intuitively synergistic, alongside less conventional yet plausible pairs such as `nautical science + civil engineering' and `optical Engineering + semiconductor engineering.'

\subsection{The CSS network analysis}
\label{sec:CCS_network}

\begin{figure}
\centering
\includegraphics[width=0.95\textwidth]{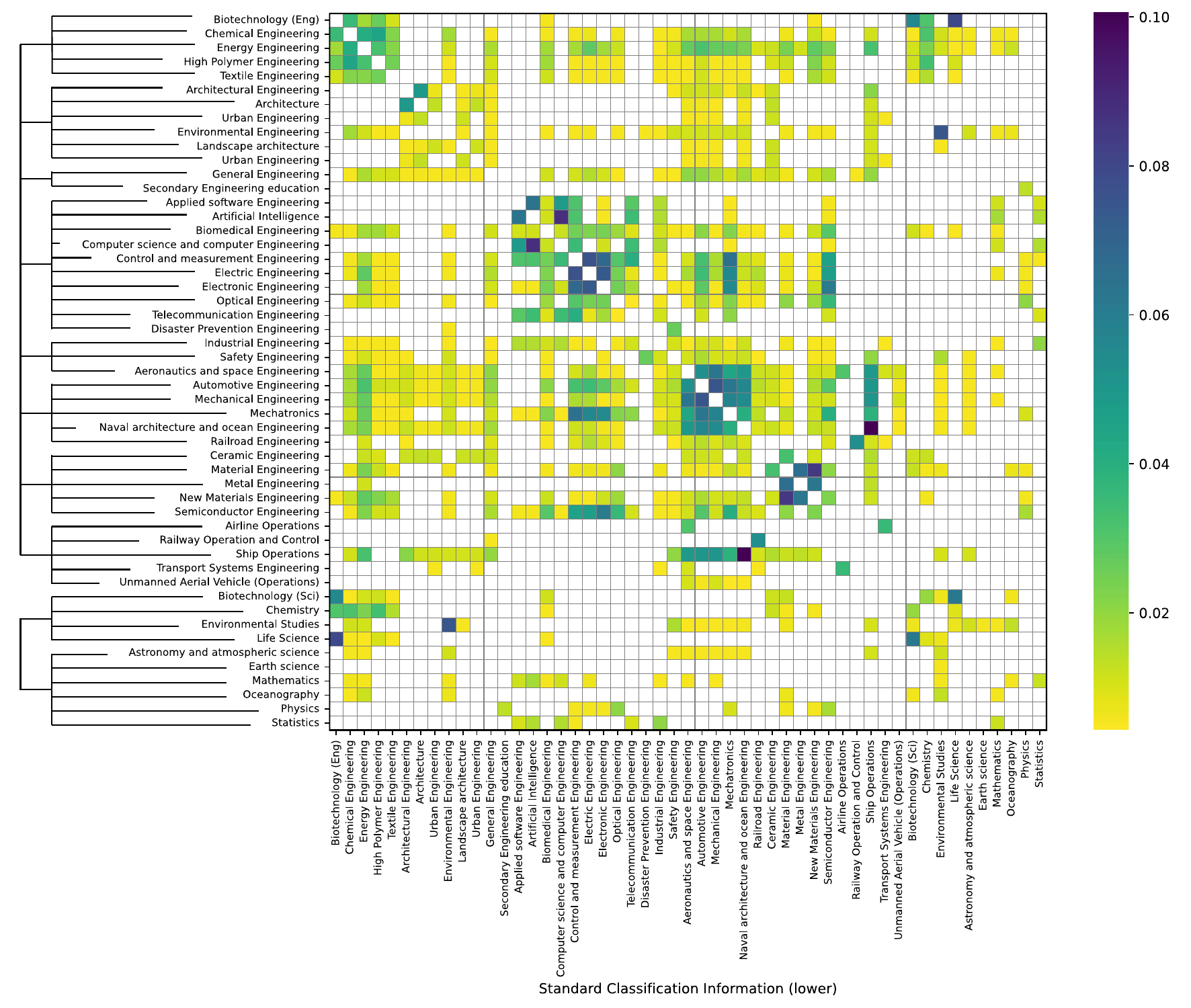} 
\caption{The heat map of CSS values between the $51$ disciplines (the lowest-level classification) in the data set. The rows and columns are sorted by the intermediate-level classification shown in the upper hierarchical level drawn on the left~\cite{data_go_kr}.}
\label{figure: heatmapt}
\end{figure}

The CSS between discipline pairs in Eq.~\eqref{eq:CSS} forms a weighted matrix or network representing the level of educational synergy between disciplines. As the number of disciplines is $51$, even a visual inspection compared to the hierarchical structure of disciplines provided by the data set~\cite{data_go_kr} is plausible. 
We show the $51 \times 51$ heatmap of the CSS for all of the STEM discipline pairs 
in Fig.~\ref{figure: heatmapt}. 
Overall, large CSS values are prominent among closely related engineering disciplines, hierarchically close to each other according to the standard classification in higher levels (forming dark block diagonal patterns)~\cite{data_go_kr}, e.g., the chemical engineering cluster, the computer science cluster, electrical engineering cluster,
the mechanical engineering cluster, and the material engineering cluster. 
In contrast, natural science disciplines do not show large CSS values to each other (the lack of dark block diagonals in the natural science region), and they have large CSS values with discipline-specific engineering counterparts, e.g., (life science $\otimes$ biotechnology) and (physics $\otimes$ semiconductor engineering). This connection pattern is an archetypal mesoscale structural example of core-periphery structures~\cite{Rombac2014}, with multiple cores composed of separate modules~\cite{Kojaku2018}.

\begin{figure}
\centering
\includegraphics[width=1\textwidth]{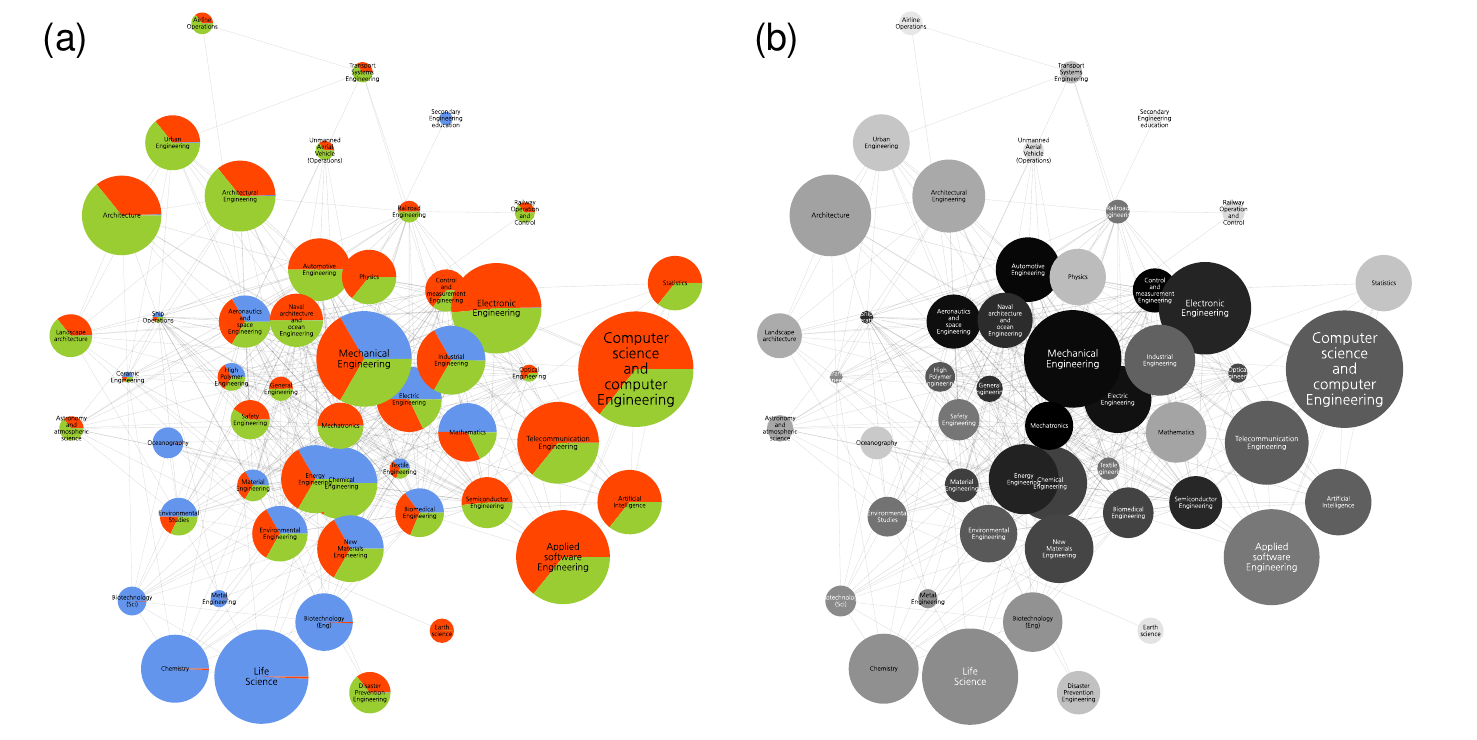} 
\caption{The community (a) and core-periphery (b) structures in the CSS network, where the size of a node (discipline) is proportional to the number of departments in the discipline. (a) The fraction of colors represents the community membership~\cite{Gupta2023} of each node, from $1000$ community ensembles with the resolution parameter $\gamma = 1$. (b) Darker colors represent larger core score~\cite{Rombac2014} values for nodes.
}
\label{figure: final networks}
\end{figure}

To characterize such a mesoscale structure, we try both community~\cite{Porter2009, Fortunato2010} and core-periphery~\cite{Rombac2014} detections. For the community structure in this discipline-synergy network, we apply the detection algorithm that allows the overlapping communities (a single node can have multiple community affiliations)~\cite{Gupta2023}, which also yields the relative strength of community membership for each node, to emphasize the interdisciplinary relations. The community structures shown in Fig.~\ref{figure: final networks}(a) indicate such overlapping communities for disciplines, in particular, drastically contrasting cases of PSE and TSE (belonging to multiple communities) and MLSE (effectively belonging to a single community).

In addition, we verify the aforementioned core-periphery separation by using  
the core score~\cite{Rombac2014}. As expected from the heat map in Fig.~\ref{figure: heatmapt}, we find that the disciplines classified as basic nature science: physics, chemistry, life science, and statistics, are located in the periphery, as shown in Fig.~\ref{figure: final networks}(b).  
In contrast, the disciplines related to engineering, which are assigned to multiple communities, show high core scores. Therefore, it verifies the existence of multiple cores~\cite{Kojaku2018} composed of engineering disciplines synergetic to each other, and the rest of the peripheral parts are composed of natural science disciplines. 

One might perceive this result as unintuitive considering the role of basic science and engineering, but in terms of academic curricula, the application side (engineering) is more synergetically attached to other disciplines than basic science, which endows most fundamental concepts and frameworks as basic resources of understanding the entire STEM fields. In other words, the interdisciplinary synergy is not readily obtained due to the unique role of natural science disciplines, which in turn suggests the necessity of maintaining their established curricula as anchoring sources of STEM education, as suggested by the CSS value of zero between physics and chemistry in Table~\ref{table:physics_chemistry}. In the case of research collaboration~\cite{HPeng2021}, of course, the pattern of interdisciplinarity can be different, so a more comprehensive understanding across education and research can be a natural candidate for future works.

\section{Discussion and conclusion}
\label{sec:conclusion}

In this paper, based on the official university-level STEM curriculum record provided by the Ministry of Education in Korea, we have proposed a comprehensive process of systematic LLM-based course standardization, the quantitative analysis of the organizational structure of departments based on the curriculum similarity, and the interdisciplinary curriculum design. We have introduced the curriculum synergy score (CSS) to evaluate the performance of the interdisciplinary curriculum model and identify synergetic discipline pairs. 

The result from the curriculum-wise organizational structure of departments suggests a deeper division between PSE and MLSE than that between science and engineering, with notable reorganization of less traditional departments depending on the organizational scale. The departments belonging to MLSE are more coherently organized, reflected by their intact community structure at finer resolution, while engineering departments show more flexible relationships within STEM disciplines. 
Such organizational flexibility of engineering departments is also highlighted in the synergetic relationship quantified by the CSS. The engineering disciplines constitute multiple core structures in the CSS network, while basic science disciplines are located in the peripheral part, with their unique sources indispensable for STEM education. 

Considering the result of our data analysis, we call for a more careful approach to designing interdisciplinary education policy. For instance, the fact that basic science disciplines provide fundamental knowledge to understand STEM does not automatically imply that it is beneficial to merge them with other disciplines; rather, we may need to acknowledge their own merit as fundamental sources of knowledge and skills. 
However, it is important to acknowledge that our algorithmic approach alone cannot overcome complex educational challenges. When our data-driven insights are integrated with evolving educational needs, we can design more effective curricula that better prepare students for future demands.
As we have demonstrated that this type of quantitative analysis can bring such insights, a natural extension of this work can be one that includes non-STEM disciplines such as literature, art, history, etc., possibly across different countries. In particular, in this era of technological breakthroughs and innovations in education, the systematic analysis of temporal change in curriculum structures is also highly desirable. On a theoretical side, the interdisciplinary curriculum models involving three or more related disciplines are a candidate for future work. We hope that our work, equipped with nationwide exhaustive data and LLM, provides a nice example for this direction.

\begin{backmatter}
\section*{Abbreviations}
STEM, science, technology, engineering, and mathematics; DSN, department similarity network; CSS, curriculum synergy score.
\section*{Declarations}
\section*{Ethics approval and consent to participate}
The data sets utilized in this study have been sourced from Higher Education in Korea (대학알리미 in Korean)~\cite{Curriculum2024,Departments2024,data_go_kr}, a public data service of the Ministry of Education in Korea. This study does not incorporate any personal information.
\section*{Consent for publication}
All of the authors consent to publication.
\section*{Availability of data and material}
The original data is available on Higher Education in Korea (대학알리미 in Korean)~\cite{Curriculum2024,Departments2024,data_go_kr}, a public data service of the Ministry of Education in Korea. The processed data, the codes, and the curriculum design are available at \url{https://github.com/Gahyoun/STEM-curriculum-2024-South-Korea-}. The full result of our optimization process in the cases of single and interdisciplinary curricula is available in the `result' in Github.
\section*{Competing interests}
The authors declare that they have no competing interests.
\section*{Funding}
This research was supported by the National Research Foundation of Korea (NRF) under the grants RS-2021-NR061247 (GG and SHL) and RS-2022-NR069864 (JY), and results of a study on the ``Gyeongsangnam-do Regional Innovation System \& Education (RISE)'' Project, supported by the Ministry of Education and Gyeongsangnam-do.
\section*{Author contributions}
GG designed the research. GG collected and processed the data. GG analyzed and visualized the data. JY provided LLM and data science technical consultation and advice. GG and SHL designed and refined the CSS. All three authors wrote the paper. All authors have read and approved the final manuscript.
\section*{Acknowledgements}
The authors thank Jibeom Seo and Beom Jun Kim for their valuable discussions and contributions during the initial stages of this research, and appreciate Daekyung Lee for providing the network-community analysis code.
\end{backmatter}

\newpage
\appendix
\section{The LLM prompts used in data processing}
\label{sec:prompts}
\begin{enumerate}
\item Tokenize the given Korean subject data to reflect the characteristics of STEM data. Example: 열역학 is about 열 and 역학, 통계물리 consists of 통계 and 물리. 전자물리 and 전자물리학 are 전자 and 물리, 화학 is 화학, 전자기학 is 전자기학, 유기생물 is about 유기 and 생물. If the data is in English, translate into Korean and discuss it.
\label{prompt1}
\item For the following list of subjects from the Korean Basic Natural Science Department, translate each subject into English and then assign each to a specific group based on the core topics and content they cover. The categorization ensures high granularity to reflect specific aspects of the subjects. Subjects like 물리학 (physics) and 물리학실험 (physics experiment) are grouped only if they address the same concepts. Closely related subjects such as 고급물리학 (advanced general physics) and 고급현대물리학 (advanced modern physics) are separated if they emphasize different principles.
\label{prompt2}
\end{enumerate}

\section{Example of tokenization and course standardization}
\label{sec:example}

We provide a step-by-step process of tokenizing and course standardization with an example of four raw course names that might be related to statistical physics. First, the following four raw course names are tokenized as follows, where $T(C_i)$ refers to the set of tokens for course $i$:
\[
\begin{aligned}
C_1 \equiv \text{Statistical Physics:} &\quad T(C_1) = \{\text{Statistics}, \text{Physics}\}, \\
C_2 \equiv \text{Statistics:} &\quad T(C_2) = \{\text{Statistics}\}, \\
C_3 \equiv \text{Thermodynamics:} &\quad T(C_3) = \{\text{Thermo}, \text{Dynamics}\}, \\
C_4 \equiv \text{Statistical Thermodynamics:} &\quad T(C_4) = \{\text{Statistics}, \text{Thermo}, \text{Dynamics}\}, \\
C_5 \equiv \text{Statistical Mechanics:} &\quad T(C_5) = \{\text{Statistics}, \text{Mechanics}\}.
\end{aligned}
\]

For each token $t_j$, we can collect the set $\mathcal{G}(t_j)$ of courses including it, for example,
\[
\mathcal{G}(\text{Statistics}) = \{C_1, C_2, C_4, C_5\}, \quad \mathcal{G}(\text{Thermo}) = \{C_3, C_4\}.
\]

After tokenization, higher-order clustering identifies semantic relationships among the courses. Using an LLM, hyperedges are formed as follows:
\begin{itemize}
\item For the token \textit{Statistics}, the LLM determines that \( C_1 \), \( C_4 \), and \( C_5 \) are semantically aligned in their scientific context, forming a hyperedge,
\[
H_1 = \{C_1, C_4, C_5\} \,.
\]

\item For the token \textit{Thermo},
\[
\mathcal{G}(\text{Thermo}) = \{C_3, C_4\} \,,
\]
and LLM groups these into another hyperedge,
\[
H_2 = \{C_3, C_4\}.
\]
\end{itemize}

Using the hyperedges \( H_1 \) and \( H_2 \), we construct a hypergraph:
\[
H_1 = \{C_1, C_4, C_5\}, \quad H_2 = \{C_3, C_4\}.
\]

From this hypergraph, connected components are identified to group courses into standardized forms. 
For the given example, the connected component is
\[
\mathcal{C}_1 = \{C_1, C_3, C_4, C_5\} \,.
\]
Therefore, the raw courses $C_1$: \textit{Statistical Physics}, $C_3$: \textit{Thermodynamics}, $C_4$: \textit{Statistical Thermodynamics}, and $C_5$: \textit{Statistical Mechanics} are grouped into a single standardized course in the context of statistical physics and thermodynamics.

\section{Community detection in appropriate resolution}
\label{sec:PaI}

\begin{figure}
\centering
\includegraphics[width=0.95\textwidth]{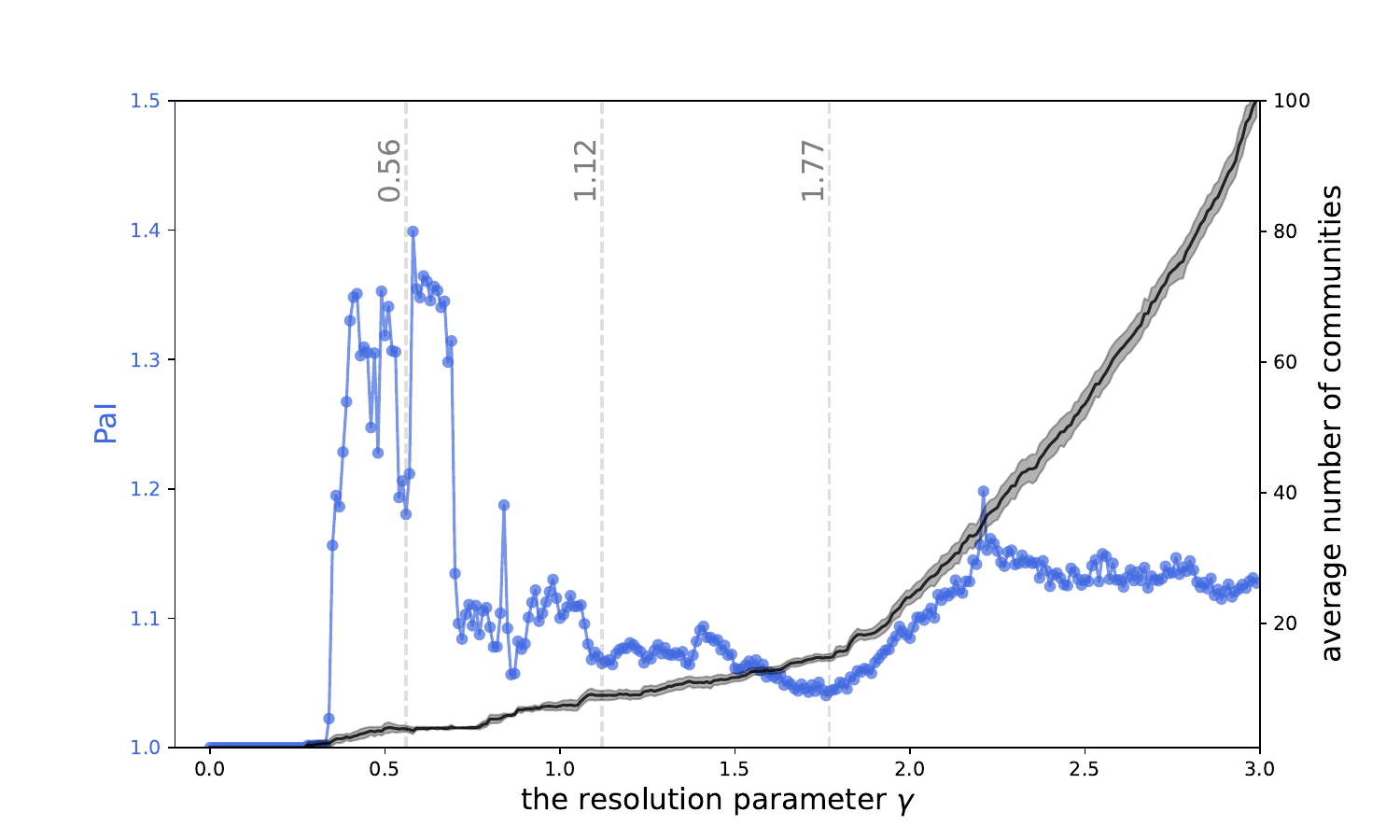} 
\caption{The PaI~\cite{PhysRevE.103.052306} values (left) and the average number of communities (right) as the functions of the resolution parameter $\gamma$ of the modularity function, as a result of $100$ independent runs of the Louvain algorithm~\cite{Blondel2008} applied to the DSN in Sect.~\ref{sec:DSN} for each $\gamma$ value. The points are the mean values, with the shades around them representing the standard deviation.}
\label{figure: PaI}
\end{figure}

The resolution parameter $\gamma$ in the modularity function~\cite{Newman2004,Reichardt2004} controls the overall scale of communities, where the modularity-maximization process with larger values of $\gamma$ tends to detect smaller communities. Although this flexibility provides a nice knob-like adjustable parameter for the scale of detected communities, it is nontrivial to decide which resolution yields the most meaningful or statistically reliable communities. In Ref.~\cite{PhysRevE.103.052306}, the authors suggest a method to extract the range of $\gamma$ corresponding to the most self-consistent detection results, namely, the regions of locally minimum values of the partition inconsistency (PaI). We apply this method to our DSN with the Louvain algorithm~\cite{Blondel2008}, and as shown in Fig.~\ref{figure: PaI}, we find there values of $\gamma = 0.56$, $1.12$, and $1.77$, corresponding to the locally consistent communities, so we take these values in addition to the trivial case $\gamma = 0$ in Sect.~\ref{sec:DSN} and Fig.~\ref{figure: department similarity network}.

\section{The log-log version of Fig.~\ref{figure: histogram and scatter}}
\label{sec:loglog_sigma}

\begin{figure}[h]
    \centering
    \includegraphics[width=0.9\textwidth]{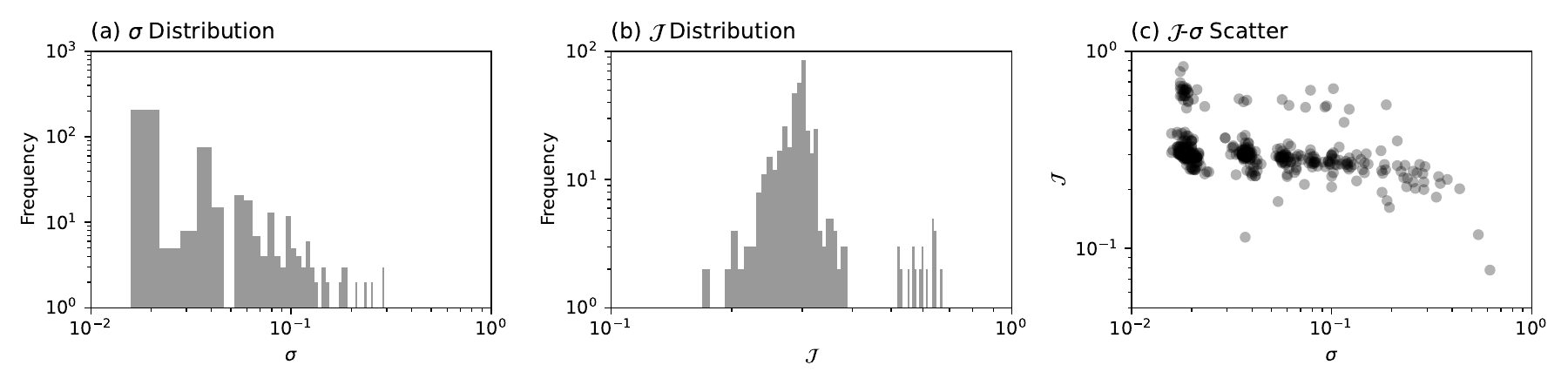}
    \caption{The log-log version of Fig.~\ref{figure: histogram and scatter}. We exclude the zero values for visualization.}
    \label{fig:loglog_sigma}
\end{figure}

To complement the linear-scale plots in Fig.~\ref{figure: histogram and scatter}, we also present the plots in the double-logarithmic scale in Fig.~\ref{fig:loglog_sigma}, to better highlight the tail behavior and sparsity of high-similarity discipline pairs.

\newpage

\begin{landscape}

\setcounter{section}{0}
\renewcommand{\thetable}{S\arabic{table}}
\renewcommand\thesection{S\arabic{section}.}

\section{The result of single and interdisciplinary curricula for `physics' and `semiconductor engineering'}
\tiny
\setlength{\tabcolsep}{5pt} 
\renewcommand{\arraystretch}{1.2} 
\begin{longtable}{p{3.5cm}|p{1cm}|p{7cm}|p{1cm}|p{7cm}|p{1cm}}
\caption{We show the curriculum design process, beginning with the seed courses at the top and adding others with the criterion described in the main text (listed in the order of addition). The courses in boldface are common in all three curricula, highlighting their central role in integrating physics and semiconductor disciplines. The CSS value for this combination is 0.02017.} \\
\hline \hline
\multicolumn{2}{c|}{\textbf{Physics Single Curriculum}} & \multicolumn{2}{c|}{\textbf{Physics \& Semiconductor Interdisciplinary Curriculum}} & \multicolumn{2}{c}{\textbf{Semiconductor Single Curriculum}} \\
\hline
\scriptsize \textbf{Course} & \scriptsize \textbf{Credits} & \scriptsize \textbf{Course} & \scriptsize \textbf{Credits} & \scriptsize \textbf{Course} & \scriptsize \textbf{Credits} \\
\hline

\scriptsize Mathematical Physics & \scriptsize 6 & \scriptsize Mathematical Physics & \scriptsize 6 & \scriptsize \textbf{Electromagnetism} & \scriptsize 6 \\
\scriptsize \textbf{Electromagnetism} & \scriptsize 6 & \scriptsize \textbf{Electromagnetism} & \scriptsize 6 & \scriptsize Electronics & \scriptsize 3 \\
\scriptsize Modern Physics & \scriptsize 3 & \scriptsize Solid State Physics & \scriptsize 3 & \scriptsize Semiconductor Manufacturing & \scriptsize 3 \\
\scriptsize Solid State Physics & \scriptsize 3 & \scriptsize Modern Physics & \scriptsize 3 & \scriptsize Circuit Theory & \scriptsize 6 \\
\scriptsize \textbf{Quantum Mechanics} & \scriptsize 6 & \scriptsize Computational Physics & \scriptsize 3 & \scriptsize Semiconductor Devices & \scriptsize 3 \\
\scriptsize Computational Physics & \scriptsize 3 & \scriptsize Electronics & \scriptsize 3 & \scriptsize Microprocessor Fundamentals and Applications & \scriptsize 3 \\
\scriptsize Mechanics & \scriptsize 6 & \scriptsize Semiconductor Manufacturing & \scriptsize 3 & \scriptsize Applied Mathematics & \scriptsize 3 \\
\scriptsize Thermal Physics and \newline Statistical Mechanics & \scriptsize 6 & \scriptsize Circuit Theory & \scriptsize 6 & \scriptsize \textbf{Semiconductor Physics} & \scriptsize 3 \\
\scriptsize Optics & \scriptsize 3 & \scriptsize Semiconductor Devices & \scriptsize 3 & \scriptsize Semiconductor Engineering & \scriptsize 3 \\
\scriptsize \textbf{Semiconductor Physics} & \scriptsize 3 & \scriptsize \textbf{Quantum Mechanics} & \scriptsize 6 & \scriptsize Semiconductor Fabrication \newline and Laboratory Practice & \scriptsize 3 \\
\scriptsize Modern Physics and \newline Experimentation & \scriptsize 3 & \scriptsize \textbf{Semiconductor Physics} & \scriptsize 3 & \scriptsize Automation and Control Systems & \scriptsize 3 \\
\scriptsize Nuclear Physics & \scriptsize 3 & \scriptsize Mechanics & \scriptsize 6 & \scriptsize Engineering Mathematics & \scriptsize 6 \\
\scriptsize General Physics and \newline Experimentation & \scriptsize 2 & \scriptsize Optics & \scriptsize 3 & \scriptsize Digital Logic and Circuit Design & \scriptsize 3 \\
\scriptsize Advanced Physics & \scriptsize 3 & \scriptsize Thermal Physics and \newline Statistical Mechanics & \scriptsize 6 & \scriptsize Semiconductor Testing and Reliability & \scriptsize 3 \\
\scriptsize Nanophysics & \scriptsize 3 & \scriptsize Nuclear Physics & \scriptsize 3 & \scriptsize Advanced Semiconductor \newline Sensor Engineering and Applications & \scriptsize 3 \\
\scriptsize Biophysics & \scriptsize 3 & \scriptsize Modern Physics and Experimentation & \scriptsize 3 & \scriptsize Display Engineering & \scriptsize 3 \\
\scriptsize Statistical Physics & \scriptsize 3 & \scriptsize Microprocessor Fundamentals and Applications & \scriptsize 3 & \scriptsize Introduction to Semiconductor Engineering & \scriptsize 3 \\
\scriptsize Wave Optics and Physics & \scriptsize 2 & \scriptsize Engineering Mathematics & \scriptsize 3 & \scriptsize Semiconductor Packaging & \scriptsize 3 \\
\scriptsize Optics and Experiments & \scriptsize 2 & \scriptsize Semiconductor Engineering & \scriptsize 3 & \scriptsize Plasma Engineering & \scriptsize 3 \\
\scriptsize Particle Physics & \scriptsize 3 & \scriptsize Introduction to Semiconductor Engineering & \scriptsize 3 & \scriptsize Semiconductor Manufacturing Processes and Equipment & \scriptsize 3 \\
\scriptsize Electronics and Physics & \scriptsize 3 & \scriptsize Semiconductor Testing and Reliability & \scriptsize 3 & \scriptsize Digital Logic Design & \scriptsize 3 \\
\scriptsize Physics Education & \scriptsize 3 & \scriptsize Advanced Physics & \scriptsize 3 & \scriptsize Analog Integrated Circuit Design & \scriptsize 3 \\
 &  & \scriptsize Automation and Control Systems & \scriptsize 3 & \scriptsize \textbf{Quantum Mechanics} & \scriptsize 3 \\
 &  &  &  & \scriptsize Artificial Intelligence & \scriptsize 3 \\
 &  &  &  & \scriptsize Digital Circuit Design \newline and Semiconductor Technology & \scriptsize 3 \\
 &  &  &  & \scriptsize Chemistry of Semiconductors & \scriptsize 3 \\
\hline\hline
\end{longtable}
\label{table:physics_semiconductor_combination}

\newpage
\section{The result of single and interdisciplinary curricula for `chemistry' and `high polymer engineering'}
\tiny
\setlength{\tabcolsep}{5pt} 
\renewcommand{\arraystretch}{1.2} 
\begin{longtable}{p{3.5cm}|p{1cm}|p{7cm}|p{1cm}|p{7cm}|p{1cm}}
\caption{We show the curriculum design process, beginning with the seed courses at the top and adding others with the criterion described in the main text (listed in the order of addition). The courses in boldface are common in all three curricula, highlighting their central role in integrating chemistry and polymer engineering disciplines. The CSS value for this combination is $0.03499$.} \\
\hline\hline
\multicolumn{2}{c|}{\textbf{Chemistry Single Curriculum}} & \multicolumn{2}{c|}{\textbf{Chemistry \& High Polymer Engineering Interdisciplinary Curriculum}} & \multicolumn{2}{c}{\textbf{High Polymer Engineering Single Curriculum}} \\
\hline
\scriptsize \textbf{Course} & \scriptsize \textbf{Credits} & \scriptsize \textbf{Course} & \scriptsize \textbf{Credits} & \scriptsize \textbf{Course} & \scriptsize \textbf{Credits} \\
\hline
\endfirsthead

\multicolumn{6}{c}{{\tablename\ \thetable{} -- continued from previous page}} \\
\hline
\multicolumn{2}{c|}{\textbf{Chemistry Single Curriculum}} & \multicolumn{2}{c|}{\textbf{Chemistry \& High Polymer Engineering Interdisciplinary Curriculum}} & \multicolumn{2}{c}{\textbf{High Polymer Engineering Single Curriculum}} \\
\hline
\scriptsize \textbf{Course} & \scriptsize \textbf{Credits} & \scriptsize \textbf{Course} & \scriptsize \textbf{Credits} & \scriptsize \textbf{Course} & \scriptsize \textbf{Credits} \\
\hline
\endhead

\hline \multicolumn{6}{|r|}{{Continued on next page}} \\ \hline
\endfoot

\hline\hline
\endlastfoot

\scriptsize
\scriptsize \textbf{Organic Chemistry} & \scriptsize 6 & \scriptsize Polymer Chemistry & \scriptsize \textbf{6} & \scriptsize Polymer Chemistry & \scriptsize \textbf{6} \\
\scriptsize Inorganic Chemistry & \scriptsize 6 & \scriptsize Polymer Properties & \scriptsize 3 & \scriptsize Polymer Properties & \scriptsize 3 \\
\scriptsize Analytical Chemistry & \scriptsize 6 & \scriptsize \textbf{Physical Chemistry} & \scriptsize \textbf{6} & \scriptsize \textbf{Physical Chemistry} & \scriptsize \textbf{6} \\
\scriptsize \textbf{Physical Chemistry} & \scriptsize \textbf{6} & \scriptsize \textbf{Organic Chemistry} & \scriptsize \textbf{6} & \scriptsize \textbf{Organic Chemistry} & \scriptsize \textbf{6} \\
\scriptsize Instrumental Analysis & \scriptsize 3 & \scriptsize Inorganic Chemistry & \scriptsize 6 & \scriptsize Polymer Characterization and Analysis & \scriptsize 3 \\
\scriptsize Biochemistry & \scriptsize 6 & \scriptsize Analytical Chemistry & \scriptsize 6 & \scriptsize Functional Polymer Materials & \scriptsize 3 \\
\scriptsize Quantum Chemistry & \scriptsize 3 & \scriptsize Instrumental Analysis & \scriptsize 3 & \scriptsize Polymer Processing & \scriptsize 3 \\
\scriptsize Chemical Kinetics & \scriptsize 3 & \scriptsize Biochemistry & \scriptsize 2 & \scriptsize Polymer Science & \scriptsize 3 \\
\scriptsize Electrochemistry & \scriptsize 3 & \scriptsize Physical Chemistry \newline and Experimentation & \scriptsize 2 & \scriptsize Polymer Composites and Design & \scriptsize 3 \\
\scriptsize Organometallic Chemistry & \scriptsize 3 & \scriptsize Polymer Composites and Design & \scriptsize 3 & \scriptsize Polymer Nano Science and Engineering & \scriptsize 3 \\
\scriptsize Physical Organic Chemistry & \scriptsize 3 & \scriptsize Polymer Processing & \scriptsize 3 & \scriptsize Chemical Engineering Stoichiometry and Process Design & \scriptsize 3 \\
\scriptsize General Chemistry \newline and Laboratory & \scriptsize 2 & \scriptsize Polymer Physics and Chemistry & \scriptsize 3 & \scriptsize Polymer Physics and Chemistry & \scriptsize 3 \\
\scriptsize Nanochemistry & \scriptsize 3 & \scriptsize Polymer Rheology & \scriptsize 3 & \scriptsize Polymer Rheology & \scriptsize 3 \\
\scriptsize Physical Chemistry and Experimentation & \scriptsize 2 & \scriptsize Functional Polymer Materials & \scriptsize 3 & \scriptsize Analytical Chemistry & \scriptsize 3 \\
\scriptsize Biochemistry Experiments & \scriptsize 2 & \scriptsize Polymer Materials for Electronics and Information & \scriptsize 3 & \scriptsize Applied Mathematics & \scriptsize 3 \\
\scriptsize & \scriptsize & \scriptsize Polymer Science and Engineering & \scriptsize 2 & \scriptsize Instrumental Analysis & \scriptsize 3 \\
\scriptsize & \scriptsize & \scriptsize Biochemistry Experiments & \scriptsize 3 & \scriptsize Polymer Materials for Electronics and Information & \scriptsize 3 \\
\scriptsize & \scriptsize & \scriptsize Polymer Science & \scriptsize 3 & \scriptsize Polymer Science and Engineering & \scriptsize 2 \\
\scriptsize & \scriptsize & \scriptsize Unit Operations & \scriptsize 3 & \scriptsize Unit Operations & \scriptsize 3 \\
\scriptsize & \scriptsize & \scriptsize Engineering Mathematics & \scriptsize 3 & \scriptsize Engineering Mathematics & \scriptsize 3 \\
\scriptsize & \scriptsize & \scriptsize Electrochemistry & \scriptsize 3 & \scriptsize Polymer Properties Experimentation & \scriptsize 2 \\
\scriptsize & \scriptsize & \scriptsize Polymer Nano Science and Engineering & \scriptsize 3 & \scriptsize Biochemistry & \scriptsize 2 \\
\scriptsize & \scriptsize & \scriptsize & \scriptsize & \scriptsize Inorganic Chemistry & \scriptsize 3 \\
\scriptsize & \scriptsize & \scriptsize & \scriptsize & \scriptsize Polymer Surface and Interface Science & \scriptsize 3 \\
\scriptsize & \scriptsize & \scriptsize & \scriptsize & \scriptsize Polymer Physics & \scriptsize 3 \\
\end{longtable}
\label{table:chemistry_high_polymer_combination}

\newpage
\section{The result of single and interdisciplinary curricula for `physics' and `chemistry'}
\tiny

\setlength{\tabcolsep}{5pt} 
\renewcommand{\arraystretch}{1.2} 
\begin{longtable}{p{3.5cm}|p{1cm}|p{7cm}|p{1cm}|p{7cm}|p{1cm}}
\caption{We show the curriculum design process, beginning with the seed courses at the top and adding others with the criterion described in the main text (listed in the order of addition). The CSS value for this combination is $0$.} \\
\hline\hline
\multicolumn{2}{c|}{\scriptsize Physics Single Curriculum} & \multicolumn{2}{c|}{\scriptsize Physics \& Chemistry Interdisciplinary Curriculum} & \multicolumn{2}{c}{\scriptsize Chemistry Single Curriculum} \\
\hline
\scriptsize Course & \scriptsize Credits & \scriptsize Course & \scriptsize Credits & \scriptsize Course & \scriptsize Credits \\
\hline
\endfirsthead
\multicolumn{6}{c}{{\tablename\ \thetable{} -- continued from previous page}} \\
\hline
\multicolumn{2}{c|}{\scriptsize Physics Single Curriculum} & \multicolumn{2}{c|}{\scriptsize Physics \& Chemistry Interdisciplinary Curriculum} & \multicolumn{2}{c}{\scriptsize Chemistry Single Curriculum} \\
\hline
\scriptsize Course & \scriptsize Credits & \scriptsize Course & \scriptsize Credits & \scriptsize Course & \scriptsize Credits \\
\hline
\endhead

\hline \multicolumn{6}{|r|}{{Continued on next page}} \\ \hline
\endfoot

\hline\hline
\endlastfoot

\scriptsize Mathematical Physics & \scriptsize 6 & \scriptsize Mathematical Physics & \scriptsize 6 & \scriptsize Organic Chemistry & \scriptsize 6 \\
\scriptsize Electromagnetism & \scriptsize 6 & \scriptsize Electromagnetism & \scriptsize 6 & \scriptsize Inorganic Chemistry & \scriptsize 6 \\
\scriptsize Modern Physics & \scriptsize 3 & \scriptsize Solid State Physics & \scriptsize 3 & \scriptsize Analytical Chemistry & \scriptsize 6 \\
\scriptsize Solid State Physics & \scriptsize 3 & \scriptsize Modern Physics & \scriptsize 3 & \scriptsize Physical Chemistry & \scriptsize 6 \\
\scriptsize Quantum Mechanics & \scriptsize 6 & \scriptsize Computational Physics & \scriptsize 3 & \scriptsize Inorganic Materials Science and Engineering & \scriptsize 3 \\
\scriptsize Computational Physics & \scriptsize 3 & \scriptsize Organic Chemistry & \scriptsize 6 & \scriptsize Instrumental Analysis & \scriptsize 3 \\
\scriptsize Mechanics & \scriptsize 6 & \scriptsize Inorganic Chemistry & \scriptsize 6 & \scriptsize Biochemistry & \scriptsize 6 \\
\scriptsize Thermal Physics and Statistical Mechanics & \scriptsize 6 & \scriptsize Physical Chemistry & \scriptsize 6 & \scriptsize Physical Chemistry and Experimentation & \scriptsize 2 \\
\scriptsize Optics & \scriptsize 3 & \scriptsize Analytical Chemistry & \scriptsize 6 & \scriptsize Polymer Chemistry & \scriptsize 3 \\
\scriptsize Semiconductor Physics & \scriptsize 3 & \scriptsize Inorganic Materials Science and Engineering & \scriptsize 3 & \scriptsize Quantum Chemistry & \scriptsize 3 \\
\scriptsize Modern Physics and Experimentation & \scriptsize 3 & \scriptsize Instrumental Analysis & \scriptsize 6 & \scriptsize Biochemistry Experiments & \scriptsize 2 \\
\scriptsize Nuclear Physics & \scriptsize 3 & \scriptsize Biochemistry & \scriptsize 6 & \scriptsize Nanochemistry & \scriptsize 3 \\
\scriptsize General Physics and Experimentation & \scriptsize 2 & \scriptsize Polymer Chemistry & \scriptsize 3 & \scriptsize General Chemistry and Laboratory & \scriptsize 2 \\
\scriptsize Advanced Physics & \scriptsize 3 & \scriptsize Quantum Mechanics & \scriptsize 6 & \scriptsize Chemical Kinetics & \scriptsize 3 \\
\scriptsize Nanophysics & \scriptsize 3 & \scriptsize Physical Chemistry and Experimentation & \scriptsize 2 & \scriptsize Electrochemistry & \scriptsize 3 \\
\scriptsize Biophysics & \scriptsize 3 & \scriptsize Mechanics & \scriptsize 6 & \scriptsize Organometallic Chemistry & \scriptsize 3 \\
\scriptsize Statistical Physics & \scriptsize 3 & \scriptsize Thermal Physics and Statistical Mechanics & \scriptsize 6 & \scriptsize Physical Organic Chemistry & \scriptsize 3 \\
\scriptsize Wave Optics and Physics & \scriptsize 3 & \scriptsize Optics & \scriptsize 3 & \scriptsize & \scriptsize \\
\scriptsize Optics and Experiments & \scriptsize 2 & \scriptsize & \scriptsize & \scriptsize & \scriptsize \\
\scriptsize Particle Physics & \scriptsize 3 & \scriptsize & \scriptsize & \scriptsize & \scriptsize \\
\scriptsize Electronics and Physics & \scriptsize 3 & \scriptsize & \scriptsize & \scriptsize & \scriptsize \\
\scriptsize Physics Education & \scriptsize 3 & \scriptsize & \scriptsize & \scriptsize & \scriptsize \\

\end{longtable}
\label{table:physics_chemistry}
\end{landscape}

\end{document}